\documentclass[fleqn,usenatbib]{mnras}

\usepackage{newtxtext,newtxmath}
\usepackage{comment}

\usepackage[T1]{fontenc}
\usepackage{ae,aecompl}

\usepackage{graphicx}	
\usepackage{amsmath}	
\usepackage{color}
\usepackage{soul}
\usepackage{ulem}

\newcommand{\argmax}{\mathop{\rm arg\ max}\limits}
\newcommand{\argmin}{\mathop{\rm arg\ min}\limits}

\newcommand{\comout}[1]{}

\title[Void and Tidal Field]{Dynamical Evolution of Voids with Surrounding Gravitational Tidal Field}

\author[M. Minoguchi et al.]{
Mutsumi Minoguchi,$^{1}$\thanks{E-mail: minoguchi.mutsumi@f.mbox.nagoya-u.ac.jp}
Atsushi J. Nishizawa,$^{2,1}$
Tsutomu T. Takeuchi$^{1,3}$ and
\newauthor
Naoshi Sugiyama$^{1,4}$
\\
$^{1}$Division of Particle and Astrophysical Science, Nagoya University, Chikusa, Nagoya 464--8602, Japan \\
$^{2}$Institute for Advanced Research, Nagoya University, Chikusa, Nagoya 464--8601 Japan \\
$^{3}$The Research Centre for Statistical Machine Learning, the Institute of Statistical Mathematics, 10--3 Midori, Tachikawa, Tokyo\\ 190--8562, Japan\\
$^{4}$Kavli Institute for the Physics and Mathematics of the Universe (Kavli IPMU, WPI), UTIAS, The University of Tokyo, Kashiwa,\\ Chiba 277--8583, Japan
}

\date{Accepted XXX. Received YYY; in original form ZZZ}

\pubyear{2015}

\begin{document}
\label{firstpage}
\pagerange{\pageref{firstpage}--\pageref{lastpage}}
\maketitle

\begin{abstract}
The void ellipticity distribution today can be well explained by the tidal field. Going a step further from the overall distribution, we investigate individuality on the tidal response of void shape in non-linear dynamical evolution. We perform an $N$-body simulation and trace individual voids using particle ID. The voids are defined based on Voronoi tessellation and watershed algorithm, using public code \texttt{VIDE}. A positive correlation is found between the time variation of void ellipticity and tidal field around a void if the void maintains its constituent particles. Such voids tend to have smaller mass densities. Conversely, not a few voids significantly deform by particle exchange, rather than the tidal field. Those voids may prevent us from correctly probing a quadrupole field of gravity out of a void shape.
\end{abstract}

\begin{keywords}
large-scale structure of Universe -- methods: numerical -- methods: data analysis
\end{keywords}


\section{Introduction}
\label{sec:introduction}
Galaxy redshift surveys have revealed that galaxies and galaxy clusters form a frothy structure called the cosmic web. One widely accepted scenario for the formation of the cosmic web currently is that the quantum fluctuations have produced the primordial density fluctuations during the epoch of inflation and the initial density fluctuations grow mainly by gravitational force at the later stages of the Universe. 
According to this scenario, the cosmic web is the promising probe of the early universe and theory of gravity.

A low-density region in the universe is called a \textit{void}. Galaxy surveys have revealed that the typical size of the void is larger than a few Mpc. In such a vast structure, gravitational force, which is a long-ranged force,  dominates comparing to others, such as electromagnetic interaction. Therefore, voids are expected to be a pure probe of gravitational theory and cosmological model. In fact, there are a number of works; voids are a promising probe of dark matter \citep[e.g.][]{ Hellwing_2009,Hamaus_2016}, dark energy \citep[e.g.][]{Lee_2009,Bos_2012,Lavaux_2012}, or gravitational theories \citep[e.g.][]{Li_2012,Cai_2015,Zivick_2015,Lam_2015}.
One of the leading difficulties of using voids for cosmology has been that we can observe fewer voids compared to galaxies. However, it has been resolved rapidly by the recent progress of the observations (e.g. SDSS \citep{Ahumada:2019aa}, HSC \citep{Aihara:2019aa}, 2dFGRS \citep{Colless:aa}, 6dFGS \citep{Jones:2004aa}) and there are still more ongoing galaxy surveys that are being planned in the near future  (e.g. LSST \citep{Tyson:2002aa}, WFIRST \citep{Green:2012aa}, Euclid \citep{Laureijs:2011aa}). 
The cosmological model can be constrained more accurately using the increased number of voids. 
On the other hand, the systematic errors {owing} to the lack of understanding of the void will become relatively critical for the correct interpretation of the results. Therefore, it becomes {increasingly} important to understand the factor that determines individual {void's} behaviour in detail.

To understand the structure formation, and build a methodology to extract information from voids, many works have been trying to model the evolution of a void. {In a simple picture,} the void is a uniform underdense region {that} expands with the background universe, growing to be more spherical under natural initial conditions \citep{Icke_1984}. 
{However, this} model is too simple to reproduce the realistic behaviour of voids in the cosmic web. 
In reality, we expect that the tidal field around void mainly affects to modify the shape of the void during cosmic history.
\cite{Park_2007} has estimated the effect of tidal force on the voids' shape distribution function by using Zel'dovich approximation, and with their model, the voids tend to be non-spherical. 
This result is also well supported by $N$-body simulation in terms of the overall distribution function; their ellipticity distribution function well fit{s} $N$-body simulation.
However, statistical properties averaged over the whole sample can also be affected by the void formation or void merger. The non-linear velocities or local structures also can change the shape, and we do not know precisely how these affect the void. 
Therefore, it is worth exploring whether the tidal field is the leading cause of {all voids'} shape evolution. 
Although the correlation between void shape and tidal field at redshift $z=0$ has also been examined by \cite{Platen_2008}, the time evolution has not been explored. 
On the other hand, \cite{Wojtak_2016} has shown that the shape of a void in the cosmic web in the $\Lambda$CDM Universe rotates and becomes distorted with time, while it does not examine the gravitational field in detail. 
We investigate the time evolution of individual voids and its relation to the gravitational force surrounding the voids. It will help us to understand the mechanism of void formation more accurately and to find appropriate statistical methods or values to reconstruct the gravitational field behind the void. 

In this {article}, we use $N$-body simulation to trace the non-linear time evolution of the voids. There are lots of ways to find \textit{a void} in the particle or galaxy distribution. For example, some count particles within regular grids or spheres to find low-density point{s} \citep{Hoyle_2002, Colberg_2005, Padilla_2005}, while they do not speciali{s}e in tracing detailed shapes. The publicly available void finder \texttt{VIDE} \citep{Sutter_2015}, which we use, is based on Voronoi tessellation density field estimation and watershed algorithm (see \ref{ssec:voidfind} for details). With these methods, we can define the edge of the void in detailed shape. The Voronoi tessellation field estimator also has the advantage that the grid size is adaptive according to particle separation, and we do not need to specify grid size by hand.

This {article} is organized as follows.
First, we introduce the details of our simulation and our definition of a void in section \ref{sec:simulation}. 
Then, in section \ref{sec:method}, we explain how we evaluate the void ellipticity and tidal field. In section \ref{sec:results}, 
we present the results for overall distribution functions in our void catalogue, the time evolution of individual shapes of voids and tidal fields around individual voids, and the correlation between tidal field and the void shape evolution. Additionally, we discuss the observational proxy for effectively identifying the isolate voids. 
The summary is given in section \ref{sec:summary}.

\section{Simulation}
\label{sec:simulation}

\subsection{Specifications}
\label{ssec:spec}
An $N$-body simulation {is performed} {using the publicly available} code \texttt{GADGET-2} \citep{Springel_2005} with Planck $\Lambda$CDM cosmology with cosmological parameters $\Omega_m=0.31$, $\Omega_\Lambda=0.69$, $\sigma_8=0.8$, $h=0.7$ \citep{Planck_2018}. The initial condition is generated by using the second Lagrange perturbation theory \citep{Crocce_2006}, and we start the simulation from $z=20$. 
The simulation box is 500 Mpc$/h$ on a side, and it contains 512$^3$ dark matter particles.
Since the largest voids found with the current galaxy survey are around 100 Mpc/$h$, 500 Mpc/$h$ box size is enough to reproduce such the largest voids. If we focus on the voids larger than 1Mpc/$h$, which is less affected by the non-linear evolution of the structure, the number of particles given here suffices the required resolution.
\subsection{Void Finding}
\label{ssec:voidfind}
{To} define the void in our simulation, we use the public code \texttt{VIDE} (The Void IDentification and Examination Toolkit) \citep{Sutter_2015}. This code is based on \texttt{ZOBOV} (ZOnes Bordering On Voidness) \citep{Neyrinck_2008}, which uses {the} Voronoi tessellation algorithm, with which we do not need to determine {a} smoothing scale by hand when we estimate the density field to define the density peaks.

Here we revisit the void finding algorithm of \texttt{VIDE}. We find voids by following steps.
\begin{enumerate}
    \item{}
    {First, we define the density field by using \texttt{ZOBOV}. 
    It computes the bisecting planes of each pair of particles and divides the entire simulation box into Voronoi cells so that each cell contains one particle. 
    The local density associated with the particle is derived as the reciprocal of the volume of each cell.}
    \item{}
    {In the next step, we find the local minimum among the density field. The Voronoi cells are grouped to form a region called a \textit{zone} which has one local minimum in it. To define the zone, the densities of adjacent cells are compared and linked to the lower density cells. If none of the neighbours has a smaller density, that cell becomes a core of the zone.}
    \item{}
    Every zone is potential void as it has a concave density profile; however, it often has {such} low-density wall that it turns out to be merely a part of a void from a visual inspection. Such low-density walls can be strongly affected by discreteness noise. Therefore, we join zones to form a \textit{void} if the zones are separated by a wall {that} has, at least at one point, lower density than the threshold which we define afterwards. The zones with higher ridge density than the threshold are regarded as voids by themselves. 
    {Although} we do not need to do, because of step (v) {below}, we can {lower the threshold value, finding the smaller voids inside the original voids. The larger voids are called \textit{parents} and the smaller voids inside are called \textit{children}}.
    \item{}
    {An effective void radius $R_v \equiv (3V/4\pi)^{1/3}$, where $V$ denotes the total volume of Voronoi cells belonging to the void, is calculated and small voids with effective radii smaller than the resolution of the simulation (here $R_v \sim 1{\rm Mpc}/h$) are excluded. Also, we exclude voids with central density higher than another density threshold to be determined to rule out Poisson noise or halos mimicking a void. Usually, \texttt{VIDE} makes cutoff using central density which is defined by particle number within {$R_v/4$}, but with this quantity, small voids which may come from Poisson noise as shown in \cite{Neyrinck_2008} is apt to be included. Therefore we use a core density $\rho_{\rm core}$, which is the reciprocal of the largest Voronoi cell in the void, as a cutoff criterion instead.}
    \item{}
    {Finally, we can optionally select void-hierarchy. We remove the voids {which are contained by any parent void } to avoid the double-counting of the ancestor voids when we trace the void in different snapshots for later analysis (see section \ref{ssec:mergertree}).}
\end{enumerate}

We summarize the fundamental quantities of voids as follows:
\renewcommand{\labelitemi}{-}
\begin{itemize}
    \item $R_v$: effective radius of the void (see (iv) above)
    \item $\rho_{\rm core}$: core density, the {reciprocal of the largest} Voronoi cell in the void
    \item $\rho_v$: void mass density, the total mass of void member particle divided by the void volume
    \item $e$: the ellipticity of the void (see following text and section \ref{ssec:ellipse})
\end{itemize}
We find that the ellipticity defined using an inertia tensor does not necessarily represent the shape of the void in the case where the dense clumps are embedded in the wall. Therefore, in this {article}, we use an alternative definition of ellipticity {introduced} in section \ref{ssec:ellipse}.

Here we have three parameters to be determined by hand. First is the zone-joint parameter in \texttt{ZOBOV} which appears in (iii) above. We set this parameter $0.2 \bar{\rho}$ where $\bar{\rho}$ is the mean matter density of the Universe. This means that our void sample does not include zones with a minimum density larger than $0.2\bar{\rho}$. The second parameter is the central density cutoff in (iv). According to \cite{Neyrinck_2008}, the voids whose core density is higher than $0.2\bar{\rho}$ are potentially affected by the Poisson noise at {$z=0$}. Therefore, here we use core density for the cutoff criterion and take this parameter as $0.2\bar{\rho}$.
The last one is void-hierarchy selection. The void-hierarchy quantifies the level of nesting of the voids; when a void {is not contained by any larger void, }void-hierarchy is 0. The higher hierarchy is recursively defined. If a void is contained by the void of the level of $i$, void-hierarchy is $i+1$. As also mentioned in (v), we take only void-hierarchy 0 voids, which means that the voids are spatially not overlapped with each other.

We have already discarded the voids potentially {arise} from Poisson noise in terms of core density, but we place further selection with the density contrast of voids to remove void-like objects from noise thoroughly. \texttt{ZOBOV} calculates the probability that a void arises from Poisson noise {by} using a fit to the probability distribution of density contrast of the voids in Poisson particle distribution (see equation (1) of \cite{Neyrinck_2008}). We remove the voids of which noise probability exceeds 5\%.

\subsection{Merger Tree of Void and Void Tracing}
\label{ssec:mergertree}
To study the time evolution of voids, we prepare two snapshots of the simulation at a given redshift {that} has slightly different cosmic time. For each of the snapshots, we employ the same method described in section \ref{ssec:voidfind}. We have tracing criteria using particle ID {to trace voids} as introduced by \cite{Sutter_2014}. We consider all pairs of a low-redshift void A (denotes 'after' evolution) and a high-redshift void B (denotes 'before' evolution) and estimate the following two quantities for all pairs: unification parameter
\begin{equation}
    UP=N_{{\rm A}\cap {\rm B}}/N_{\rm A}
\end{equation}
and division parameter 
\begin{equation}
    DP=N_{{\rm A}\cap {\rm B}}/N_{\rm B},
\end{equation}
where $N_{\rm A}$ and $N_{\rm B}$ are the numbers of constituent particles for void A and B, respectively, and $N_{{\rm A}\cap {\rm B}}$ denotes the number of particles shared by both void A and B. Both the two parameters are indicators of particle retention but independent of each other. $UP$ takes the maximum value of 1 if all member particles of descendant void A come from ancestor void B. On the other hand, if void A inherits all the member particles of void B, then $DP$ takes the value of 1. We then consider {that} void A and void B are identical only when both $UP$ and $DP$ {are} sufficiently high, where significant {mergers or divisions do not occur} during their time evolution. 

We first calculate $UP$ and $DP$ for all pairs of voids. Then for given descendant void A$_i$, we define the candidate ancestor void B which maximizes the $UP$. Conversely, we also find the best candidate for {the} given void B$_j$ by looking at $DP$. We connect void A and B only when the best candidates coincide with each other.

To quantify how many particles remain in the void during evolution, we define particle retention
\begin{equation}
    PR=\sqrt{UP\ DP}.
\end{equation}
When $PR$ is high, it means that the void retains member particles. Therefore, we can focus only on the voids which are less affected by merger or division by looking at the voids with high $PR$. 

\begin{figure}
  \centering
  \includegraphics[width=8cm]{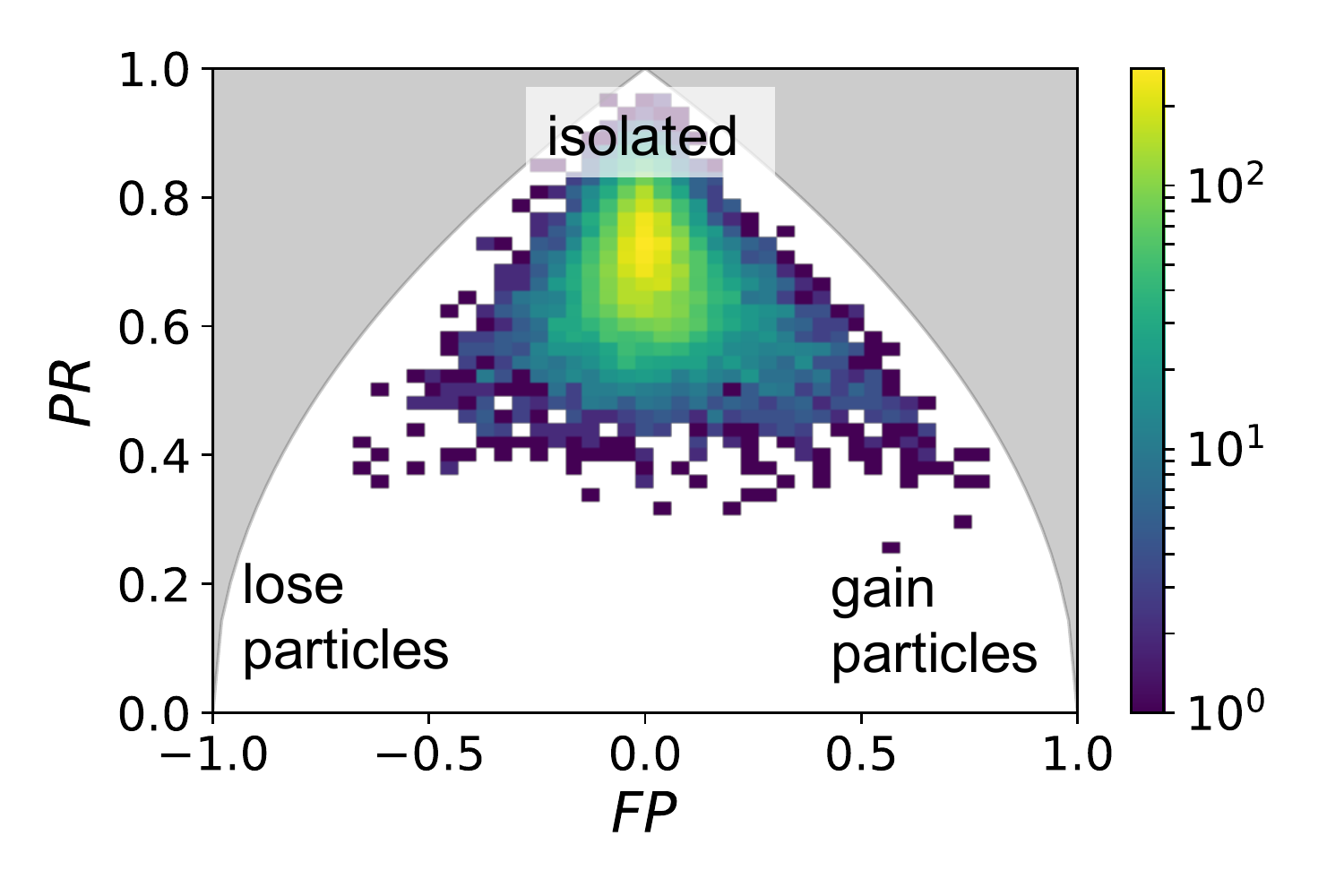}
  \caption{The number distribution of voids as a function of $FP$ and $PR$. The definition of parameters {forbids the grey regions}. The voids with high $PR$ (marked 'isolated' in the figure) exchanges very few particles compared to the number of its member particles. If $FP$ is high ('gain particles' in the figure), a void gain particles from outside of its progenitor, and if it is low ('lose particles' in the figure), the majority of the member particles flow out of the descendant void.
  }
\label{fig: PR}
\end{figure}

Also, by looking at the flow parameter
\begin{equation}
    FP=\frac{N_{\rm A}-N_{\rm B}}{N_{\rm A}+N_{\rm B}-N_{{\rm A}\cap {\rm B}}},
\end{equation}
we can further distinguish whether particles immigrate from other voids or emigrate to others. 
As is shown in Figure \ref{fig: PR}, high $PR$ corresponds to $FP\sim 0$, as no particle exchange occurs.

In this {article}, we calculate time derivatives of fundamental quantities of voids at $z \sim 0$.
For this purpose, we take a time interval sufficiently shorter than the typical time scale of the void evolution, which can be roughly estimated as follows. In our catalogue, the median of void mass density/core density is $\rho_v\sim1.5\bar\rho$ or $\rho_{\rm core}\sim0.1\bar\rho$, where $\bar{\rho}$ is an average mass density of the Universe. Then {the} dynamical timescale of a void is roughly $t\sim(G\rho_v)^{-1/2}\sim60$ Gyr. Therefore, the timestep of $1$ Gyr should be reasonable to trace the dynamical evolution of the voids. In practical N-body simulation, 1 Gyr at $z=0$ takes about 100 timesteps which also seems reasonable for smooth particle motions. 
For later convenience, 
we introduce the notation for the time evolution of physical quantity $X$ as 
\begin{equation}
    \Delta X\equiv X(t=t_1)-X(t=t_0),
\end{equation}
where $t_0$ is the time taken at $z=0$ and $t_1$ is the time 1 Gyr after $t_0$.

\begin{figure}
  \centering
  \includegraphics[width=8cm]{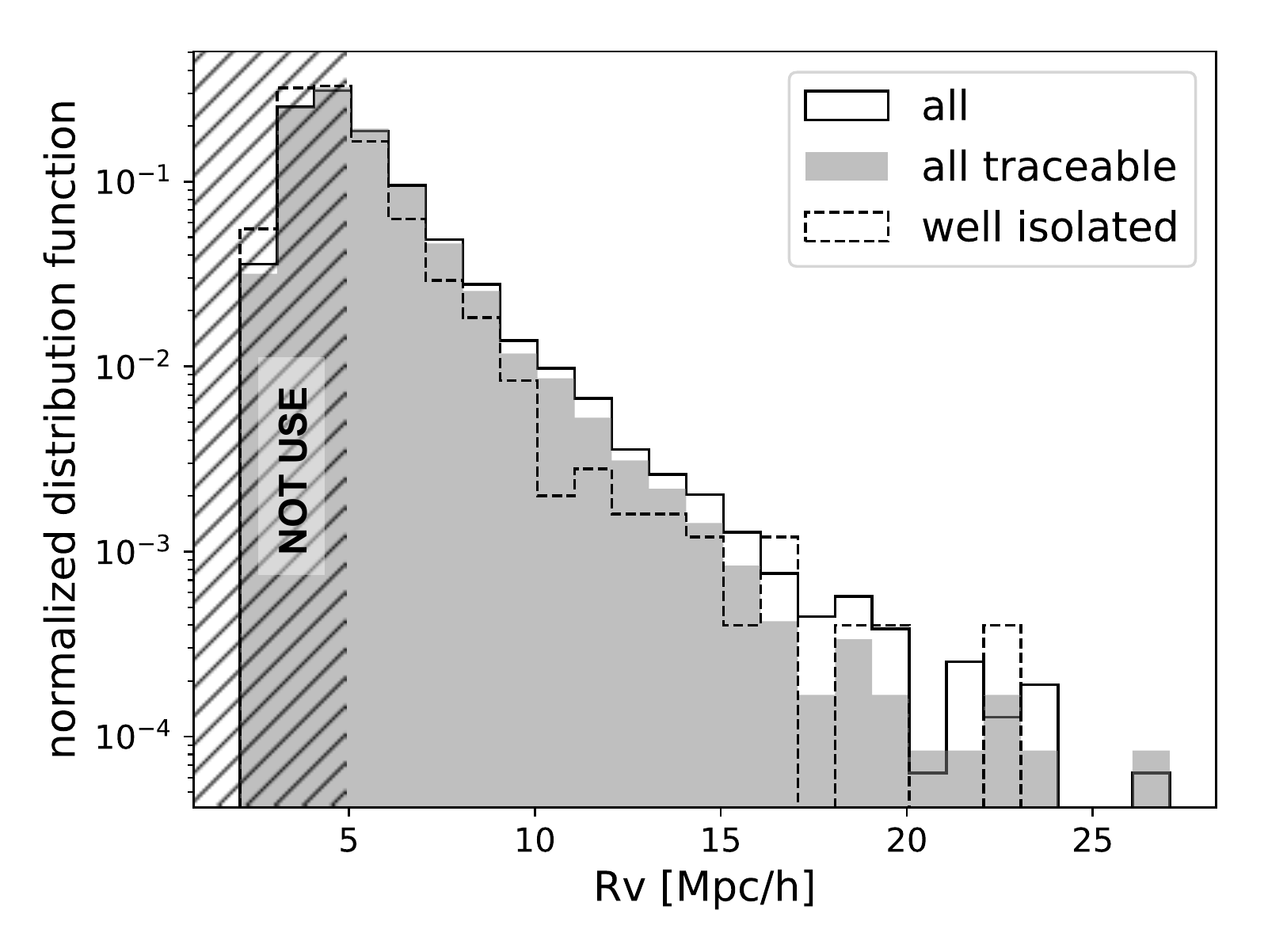}
  \caption{The probability density functions of void sizes at $z=0$ in each criterion. 'all traceable' means that the voids do not die within 1 Gyr and 'well isolated' denotes the voids with $PR>0.75$, which means that the voids exchange fewer particles in the process of evolution. {We do not use the voids smaller than 5Mpc/h in size in the following analysis to avoid the resolution effect.}
  }
\label{fig: R_n}
\end{figure}

Finally, {the probability density functions of void sizes are shown} in Figure \ref{fig: R_n}. There are 58457 voids (indicated by 'all' in the legend) found by following (i)-(v) in section \ref{ssec:voidfind}, while 11915 of them are traceable (indicated by 'all traceable' in the legend). The size distributions of these two are almost the same except small difference at large $R_v$. We additionally show the distribution of the 2504 'well isolated' voids, whose $PR$ is higher than 0.75 (exchange fewer particles during the evolution). Again, the distribution does not change significantly, but voids tend to be slightly smaller in this case. 
{The size distribution function has a peak at 5Mpc/h in our simulation but for the higher resolution simulation, the number of voids smaller than 5Mpc/h increases. However, we see that the number of voids larger than 5Mpc/h does not change significantly and thus we use the voids $R_v\ge5$Mpc/h hereafter to avoid the resolution effect.}

\section{Method}
\label{sec:method}
\subsection{Void Ellipticity}
\label{ssec:ellipse}
Although it is still an open question to characterize the shape of voids and various definitions have been proposed in the literature \citep[e.g.][]{Neyrinck_2008, Lavaux_2012}, we describe the void with a triaxial ellipsoid shell with axis lengths, $a_{3}\le a_{2}\le a_{1}$.
Then the ellipticity is defined as
\begin{equation}
e=1-\frac{\sqrt{a_{2}a_{3}}}{a_{1}}. 
\label{Eq: ellip}
\end{equation}
The axis lengths are given by fitting the ellipsoid to voids. The distance from the centre to the surface to the triaxial ellipsoid can be described by three axial lengths and three Y-Z-Y Euler angles, $\alpha, \beta$ and $\gamma$, 
\begin{align}
&r(\psi,\phi)=\nonumber \\
&\left[\left(\frac{\gamma_c (\beta_s \psi_s \phi_s+\beta_c (\psi_c \alpha_c \phi_s-\phi_c \alpha_s))
    -\gamma_s (\alpha_c \phi_c+\psi_c \alpha_s \phi_s)}{a_1}\right)^2 \right. \nonumber \\
&\left. +\left(\frac{\beta_c \psi_s \phi_s+\beta_s (\phi_c \alpha_s-\psi_c \alpha_c \phi_s)}{a_2}\right)^2 \right. \nonumber \\
&\left. +\left(\frac{\gamma_c (\alpha_c \phi_c+\psi_c \alpha_s \phi_s)
    +\gamma_s (\beta_s \psi_s \phi_s+\beta_c (\psi_c \alpha_c \phi_s-\phi_c \alpha_s))}{a_3}\right)^2\right]^{-1/2}, \label{ellipr}
\end{align}
where subscript $s$ and $c$ stand for the sine and cosine functions, i.e. $\phi_c \equiv \cos(\phi)$.
The axis directions are {described by the unit vectors $\mbox{\boldmath $  A  $}_i$ whose subscript $i=(1,2,3)$ corresponds to the axis length index. They are given by}
\begin{equation}
\left(
\mbox{\boldmath $  A  $}_1,
\mbox{\boldmath $  A  $}_2,
\mbox{\boldmath $  A  $}_3
\right)=
R_y(\alpha)R_z(\beta)R_y(\gamma)\left(
\mbox{\boldmath $  e  $}_x,
\mbox{\boldmath $  e  $}_y,
\mbox{\boldmath $  e  $}_z
\right).
\end{equation}
{Here} $\mbox{\boldmath $  e  $}_i\ (i=\{x,y,z\})$ are bases of global coordinates of the simulation box, and $R$ is the rotation matrix. 
To fit an ellipsoid to voids, we take a standard chi-square minimization for all the constituent particles for each void at the position $(\psi, \phi)$,
\begin{equation}
    \chi^2 = 
    \sum_{i \in {\rm void}}
    [r(\phi_i, \psi_i ) - {\mathcal{A}}R_i]^2,
\end{equation}
where $R_i$ is the {measured} distance from the void centre to each constituent particle, and ${\mathcal{A}}$ is introduced to absorb a similarity transformation. Here the void centre is defined as the average of member particle positions weighted by the Voronoi cell volumes.

\subsection{Tidal field and Void-Tide Alignment}
\label{ssec:alignment}
In this section, we describe how we measure the gravitational tidal field around voids. We take an arbitrary direction $\hat{\boldsymbol n}$ and expand the radial component of gravitational force on the spherical shell of radius $r$, centred at the centre of gravity of a void, in Legendre series;
\begin{equation}
F_{\hat{\boldsymbol n}}^{(l)}(r)=-\frac{2l+1}{2}\int^1_{-1}\frac{\partial \Phi(r, \hat{\boldsymbol \theta})}{\partial r}P_l(\mu)\ \rm{d}\mu, \label{eq_Fr}
\end{equation}
where $\mu = \hat{\boldsymbol \theta}\cdot\hat{\boldsymbol n}$ and $\Phi$ is {the} gravitational potential. We use Gadget-2 to estimate the gravitational potential. By embedding dummy massless particles, we have Gadget-2 calculate potential $\Phi$ at 3072 \texttt{Healpix} (Hierarchical Equal Area isoLatitude Pixelization) \citep{Gorski_2005} grid points on two concentric spherical shells around each void, whose interval is 1~Mpc/h. These pixels have equal solid angle $\simeq 13.5\ {\rm [{deg}]}^2$. Then we calculate the radial gradient in equation \eqref{eq_Fr} numerically. The integration in equation \eqref{eq_Fr} is approximated by the summation on Healpix grid points. We locate the direction of $\hat{\boldsymbol n}$ where $F^{(l)}$ is maximized or minimized; the direction that maximizes or minimizes the $l$-th multipole mode is written as
\begin{equation}
\hat{\boldsymbol n}_{\max}^{(l)}{(r)}=\argmax_{\hat{\boldsymbol n}}\left[ F_{\hat{\boldsymbol n}}^{(l)}{(r)} \right],\ \ \ \ \ 
\hat{\boldsymbol n}_{\min}^{(l)}{(r)}=\argmin_{\hat{\boldsymbol n}}\left[ F_{\hat{\boldsymbol n}}^{(l)}{(r)} \right],
\end{equation}
and the $F^{(l)}$ expanded in those coordinates can be denoted as
\begin{equation}
F_{\max}^{(l)}(r)=\max_{\hat{\boldsymbol n}}\left[F_{\hat{\boldsymbol n}}^{(l)}(r) \right],\ \ \ \ \ 
F_{\min}^{(l)}(r)=\min_{\hat{\boldsymbol n}}\left[F_{\hat{\boldsymbol n}}^{(l)}(r) \right].
\end{equation}
The maximum and minimum directions
$\hat{\boldsymbol n}_{\rm max}$ and
$\hat{\boldsymbol n}_{\rm min}$ are defined as the central position of the pixel. {This article focuses} on the quadratic component since ellipticity is also a quadratic approximation of void shape. Hereafter we simply refer to the quadrupole moment of radial gravitational force $F^{(2)}$ as a tidal field.

In section \ref{ssec:ellipse}, we determine void axes direction by fitting an ellipsoid to each void. 
Using the best-fitting parameters for two different snapshots, we define time variation of void major axis direction

\begin{equation}
    \vartheta_{\Delta {\rm void}}=\cos^{-1}(\mbox{\boldmath $  A  $}_{1}{(t_0)\cdot\mbox{\boldmath $  A  $}_{1}(t_1)}).
\end{equation}
along with time variation of tidal direction
\begin{equation}
    \vartheta_{\Delta {\rm tidal}}=\cos^{-1}(\hat{\boldsymbol n}^{(2)}_{\max}(R_v{,t_0)\cdot\hat{\boldsymbol n}^{(2)}_{\max}(R_v,t_1))},
\end{equation}
and void-tidal alignment
\begin{equation}
\theta(t) =\cos^{-1} [\mbox{\boldmath $A$}_1(t) \cdot \hat{\boldsymbol n}^{(2)}_{\max}(R_v,t_{0})].
\end{equation}
{Again, $t_0$ 
is the time taken at $z=0$ and $t_1$ is the time 1 Gyr after $t_0$ and we define the time variation of $\theta$ as $\Delta\theta=\theta(t_1)-\theta(t_0)$}. 
Finally, we define maximal tidal strength as
\begin{equation}
T=F_{\max}^{(2)}(R_v)
\end{equation}
and its vector component of the major axis direction of void:
\begin{equation}
T_\theta=T\cos\theta.
\end{equation}

\section{Results and Discussions}
\label{sec:results}

\subsection{Time Dependence of Averaged Properties}
\label{ssec:voidcatalogue}
In this section, we summarize the overall properties of our void catalogue {on} a cosmological time scale, from $z=1$ to $z=0$. Our void catalogue has 58458 voids at $z=0$ and 151718 voids at $z=1$. The discussions in this section are based on the comparison between averaged statistics at different epochs, without tracing the evolution of individual void.\\

\noindent
\textbf{Size}\\
At both redshifts, their sizes range from a few Mpc/$h$ to 20 Mpc/$h$ in a comoving scale. 
In our analysis, we exclude the parent voids, which include smaller children voids inside. It will help avoid the spatial overlapping of voids and enable us to define a unique void tree. After removing all parent voids, we see that the average void size is almost identical from $z=1$ to $0$: $5.6 $ and $5.2$ Mpc/$h$, respectively.
\\

\begin{figure}
\begin{center}
  \includegraphics[width=6cm]{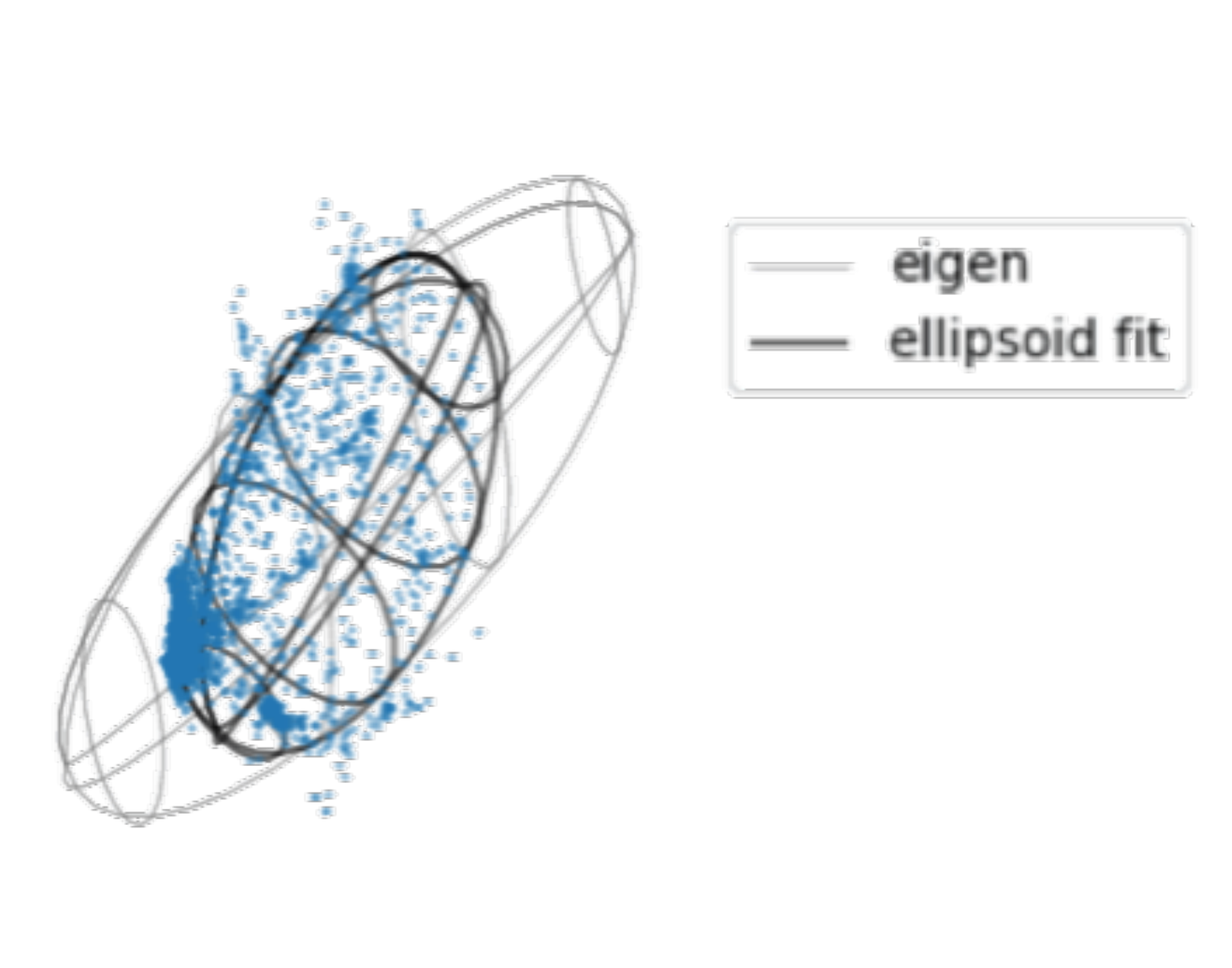}
  \end{center}
  \caption{
  Comparison of two methods for approximating void shape with ellipsoid. The dots represent the constituent dark matter particles of the void. The grey grid represents the ellipsoid characterized by the eigenvalues of inertia tensor, while the black grid represents the ellipsoid by shell fitting (see text for more detail).
  }
\label{fig: ellipfit}
\end{figure}

\noindent
\textbf{Shape}\\
\texttt{VIDE} computes the ellipticity based on the inertia tensor of the member particles of each void. Assuming that the density is uniform inside the void, we can uniquely define an ellipsoid which represents the void shape, as shown in Figure \ref{fig: ellipfit} labelled 'eigen'. However, we find the inertia tensor is strongly affected by local structures and does not necessarily represent the shape of the void. Therefore, the void shape {is determined} by ellipsoidal shell fitting described in section \ref{ssec:ellipse}. Figure \ref{fig: ellipfit} demonstrates the dark matter distribution {and} the 3-dimensional ellipsoids defined by {two different} methods, which clearly show{s} that our approach better represents the apparent shape of the void. The ellipsoid defined with the inertia tensor can reproduce the underlying dark matter distribution when dark matter is distributed almost uniformly on the ellipsoidal shell. However, as {shown} in Figure \ref{fig: ellipfit}, the dark matter is significantly localized, and ellipsoid defined in this way is different from the actual distribution of dark matter. Conversely, our method, fitting the shape of dark matter distribution with ellipsoid, can better reproduce the dark matter distribution around the void (see section \ref{ssec:ellipse}).

With this fitted ellipsoid, we compute the ellipticity defined by equation \ref{Eq: ellip}, and averaged overall voids, the mean ellipticity is $\bar{e}=0.37$ at $z=1$ and $\bar{e}=0.41$ at $z=0$, respectively.

It is in general difficult to compare the statistics of voids in the literature mainly because the different authors use a different definition of void and ellipticity. Here we compare our results of the mean ellipticity change at two different redshifts to the previous work by isolating the differences one by one. 

The reference work to be compared is \cite{Bos_2012} which has reported that the mean ellipticities at $z=1$ and $0$ are $0.45$ and $0.46$, respectively.
First, the N-body simulation data {is generated} with the same cosmological parameters and simulation specification parameters, such as box size or the number of particles. \cite{Bos_2012} use the \texttt{WVF} (Watershed Void Finder) \citep{Platen:2007aa} to find the voids while we use the \texttt{VIDE}; both methods are based on the tessellation of the particles. It is known that those two different finders show a similar shape of a void \citep{Colberg:2008aa}. Since the smoothing is not implemented in \texttt{VIDE}, we reproduce the smoothing method implemented as in \cite{Bos_2012}. The density field is smoothed on the regular grid and interpolated to every particle positions. When we compute the ellipticity, we first use the shape from shell fitting while \cite{Bos_2012} use the different definition based on the inertia tensor of the voxels (regular grid cells) inside the void. 
Then the mean ellipticities are $0.39$ and $0.43$, respectively. Once we use the same definition of ellipticity, we find $0.45$ and $0.46$, the same values as the reference work. {These facts show} that as long as we apply the density field smoothing, the {watershed-based void finders seem to generate} similar results. At the same time, we also see that the {difference in the} definition of void shape largely affects the ellipticity in this case. 
However, we note that the ellipticity {difference} does not exceed a possible systematic error from simulation resolution $\sim 1 {\rm Mpc}/h$, which can be read as 20\% for ellipticity $\sim 0.5$, with the typical void size $\lesssim$ 10 Mpc/h.
\\

\noindent
\textbf{Void-Tidal Alignment}\\
Finally, we estimate the alignment angle $\theta$ between the void major axis and tidal axis. If these two axes are
randomly rotating, $\theta$ approximately becomes 60 degrees on average. With our void catalogue, the mean alignment angle is 29 deg at $z=1$ and 26 deg at $z=0$, which decreases 0.4 deg/Gyr on average. If we define the void axis directions by eigenvectors of the inertia tensor, we obtain a slightly higher value of the angle, 29 deg at $z=0$. 

\subsection{Time Evolution of Individual Voids at \texorpdfstring{$z=0$}{z=0}}
\label{ssec:Time Evolution of individual voids at z=0}

Hereafter, we focus on the individual void evolution. We only consider traceable voids, which means that we can define the identical void in both snapshots before and after the time evolution {based} on the conditions defined in \ref{ssec:mergertree}.

\subsubsection{Rotation of Void Major Axis and Tidal Field}
\label{ssec:Rotation of Void major axis and tidal field}

In the Newtonian gravity, the gravitational potential $ \Phi $ and the density fluctuation $ \delta $ in the Fourier space are related as $ \Phi \propto k^{-2} \delta $. Therefore, we can naively expect that a smaller scale structure with stronger non-linearity in the potential field is relatively smoothed out compared to the density field. {Conversely}, the geometrical structures in the potential field, such as the peaks or saddle points, are expected to be more stable than those of the density field because of {their} relatively stronger linearity.

\begin{figure}
  \centering
  \includegraphics[width=8cm]{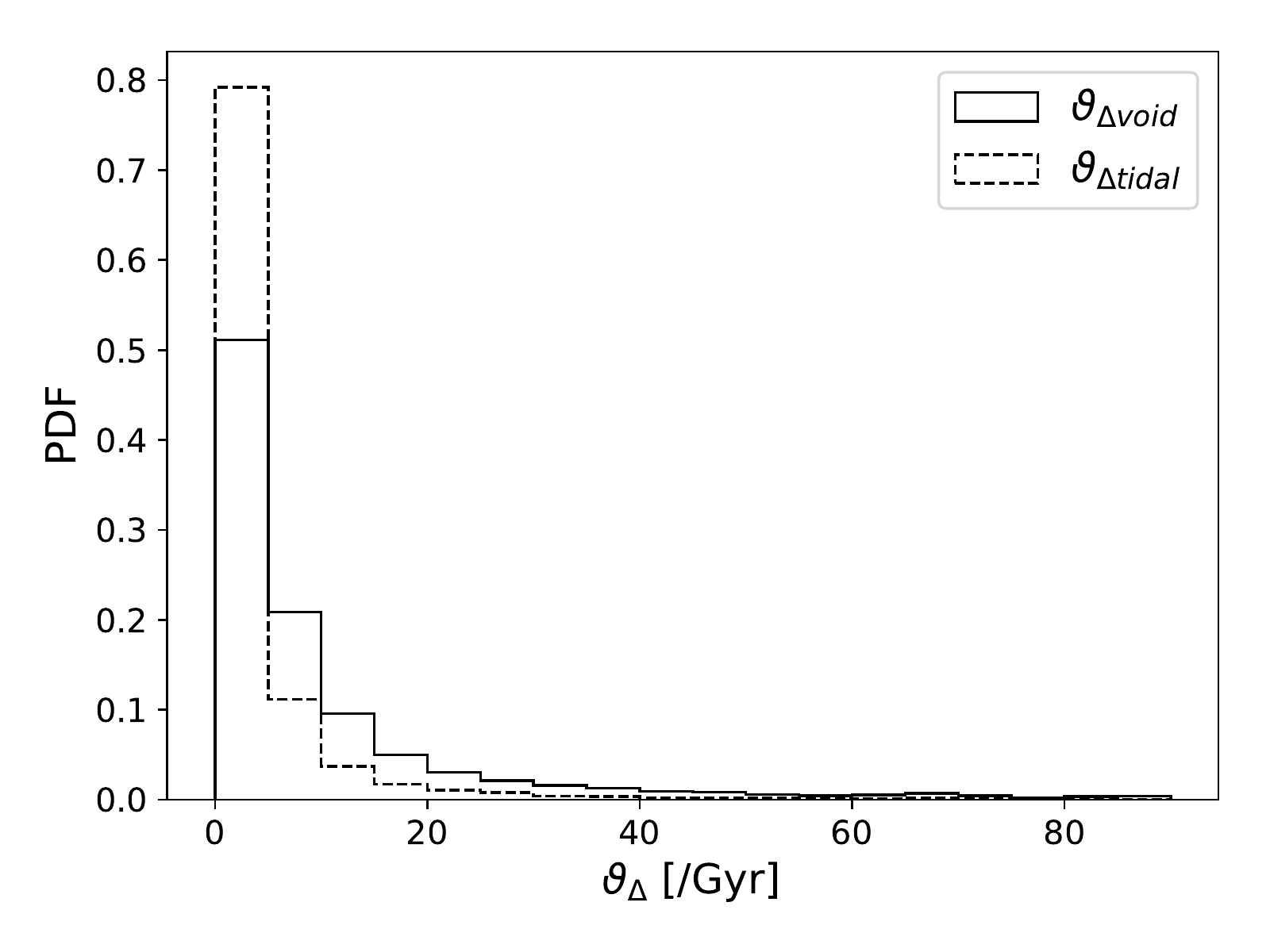}
  \caption{
  The angular variation of the void major axis directions (solid lines) or tidal quadrupole (dashed lines) in 1~Gyr for \textit{all traceable} voids. The bin width on the x-axis is 5 deg. About 80~\% of the tide field rotates only 5~deg or less, while {about a half} of voids rotate more than 5~deg.
  }
\label{fig: R_ang}
\end{figure}

Therefore, we suppose that the void rotates towards the direction of the quadrupole component of the almost fixed tidal field.
To evaluate this assumption, we examine the rotation angles at 1~Gyr concerning the major axes of the void and tidal field directions. The result {is shown} in Figure \ref{fig: R_ang}; the solid-line histogram represents the rotation angle of the major axis of the void, and the dashed-line histogram represents the rotation angle of the tidal field. In this figure, the fraction of $ \vartheta_{\Delta {\rm tidal}}$ less than our angular grid resolution, $\sim$ 5 deg, account for 78~\% of all traceable case, while {about a half} of the voids rotate more than 5 degrees. This result roughly supports our assumption.

\subsubsection{Void-Tidal Alignment}
\label{ssec:Void-Tidal Alignment}

If the void is distorted by the tidal field, the major axis of the void will become aligned with the direction of the tide field as it grows, and therefore $ \Delta \theta <0 $ is expected.

\begin{figure}
  \centering
  \includegraphics[width=8cm]{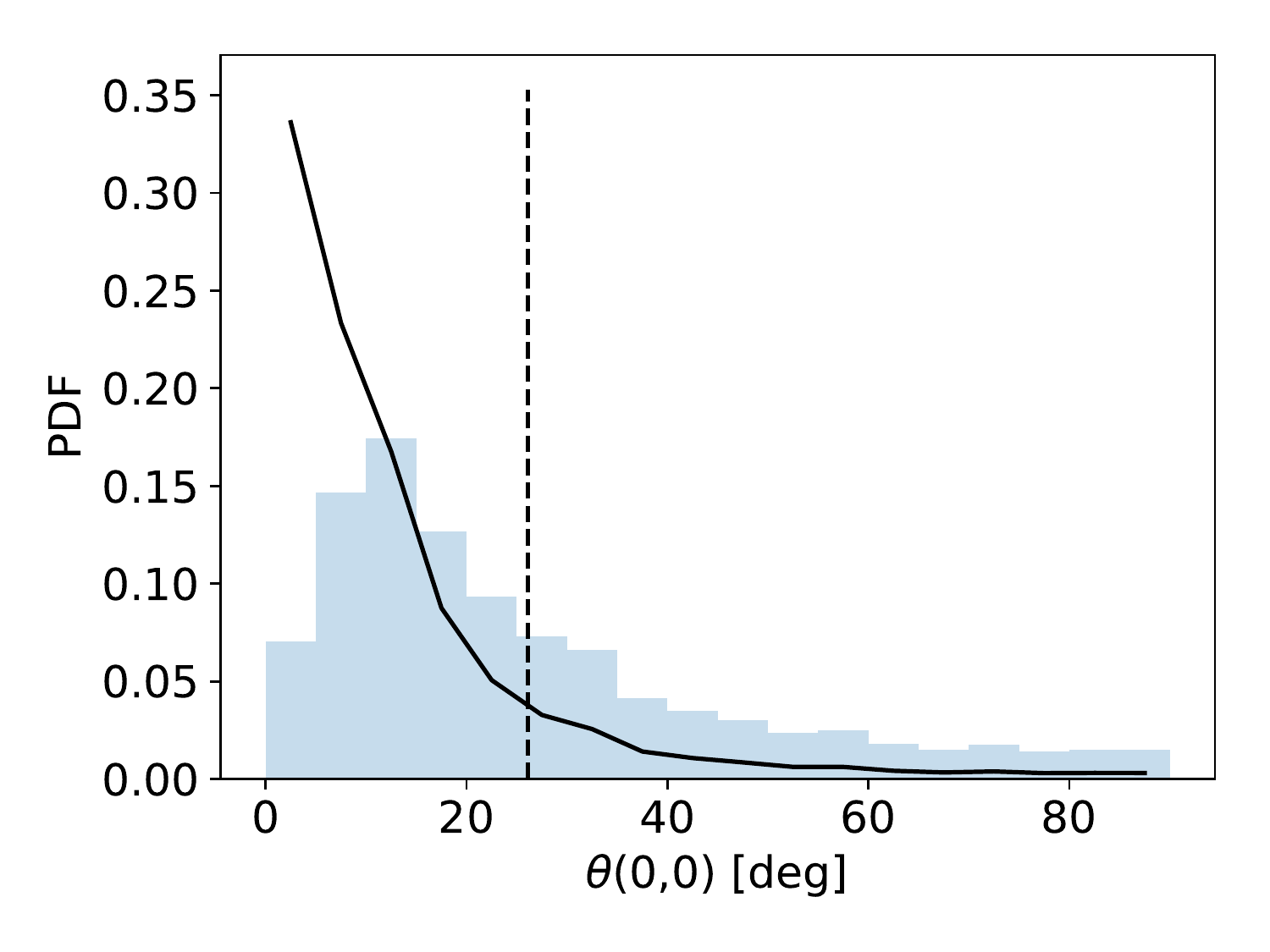}
  \caption{
  The histogram shows probability density function $P(\theta)$ for the alignment angle $\theta$ of \textit{all traceable} voids at $z=0$.
  The black and solid line is $P_a(\theta)$: the probability density function normalized by solid angle. {The vertical dashed line indicate{s} the mean value {$\theta\sim 26$}, while 60 is expected for random rotation.}
  }
\label{fig: t_at}
\end{figure}

First, we show the probability density distribution function of alignment $ \theta $ at $z=0$ in Figure \ref{fig: t_at}.
The histogram in the figure is the probability density function of alignment $ P (\theta) $, and the vertical dotted line is the average. The black line is the angular probability density function $ P_a (\theta) $, which is obtained by dividing $ P (\theta) $ by the solid angle of each bin:
{
\begin{equation}
    P(\theta_i)\ \Delta \theta_i = P_a(\theta_i)\ 2\pi \sin \theta_i \Delta \theta_i, \label{P_theta}
\end{equation}
}
{where $\theta_i$ is the middle point of the $i$-th $\theta$ bin, and $\Delta \theta_i$ is the bin width. Both $P(\theta)$ and $P_a(\theta)$ are normalized with respect to the total number of voids in Figure \ref{fig: t_at}.}
$P_a(\theta)$ is the probability where the volume effect is removed, and it takes the maximum value at $ \theta = 0 $ as shown in the figure.
The average value of the alignment of all traceable voids {is 26 deg, which is consistent with what we obtained for the overall average in section} \ref{ssec:voidcatalogue}.

\begin{figure}
  \centering
  \includegraphics[width=8cm]{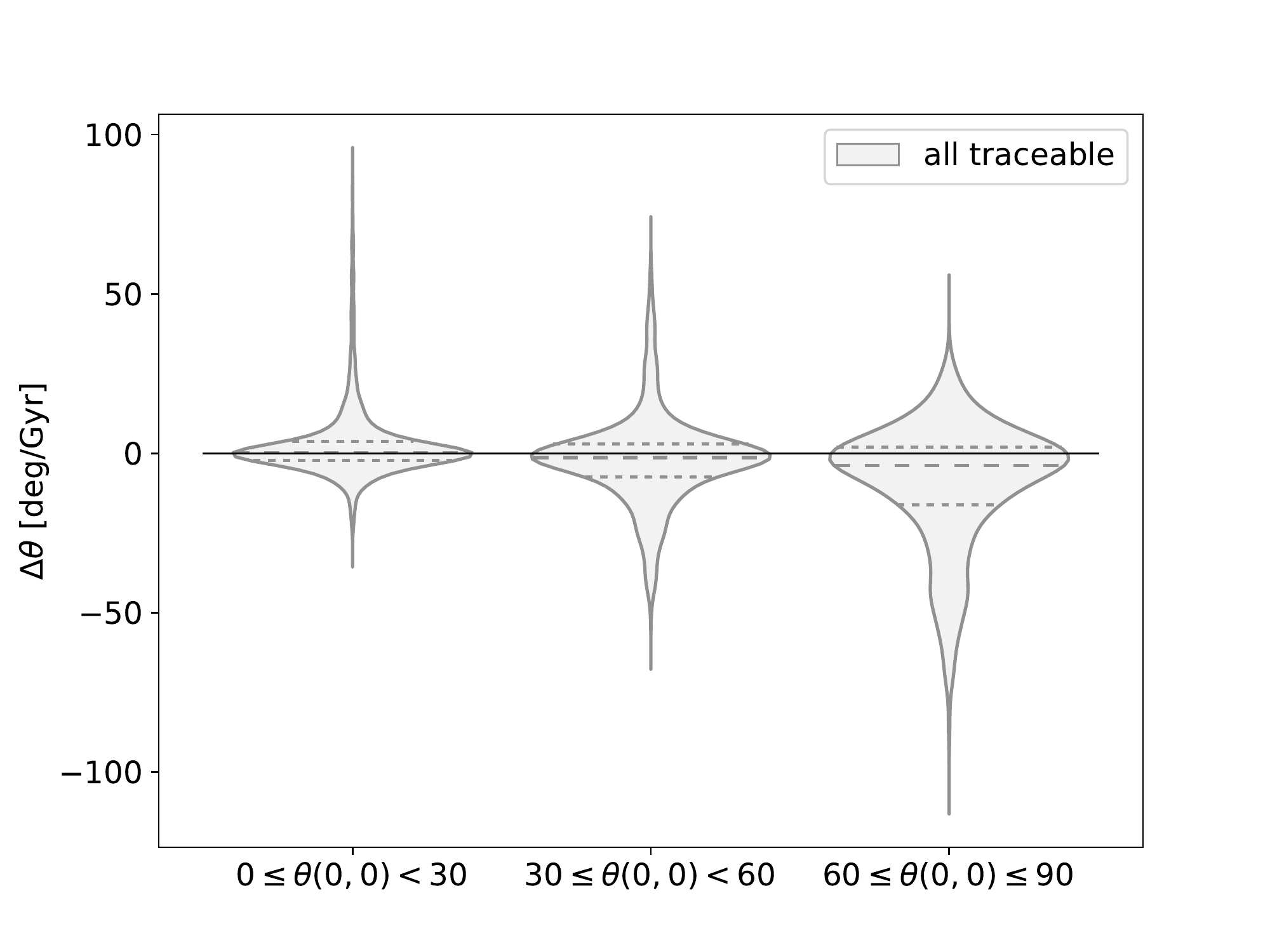}
  \caption{
  The distributions of alignment increase per Gyr at $z = 0$. The dashed line in each violin plot represents the median, and the dotted lines represent 25 and 75 percentiles.}
\label{fig: t_dt}
\end{figure}

On the other hand, Figure \ref{fig: t_dt} shows the time evolution tendency of $\theta$. The vertical axis in the figure is the amount of change of $\theta$ per Gyr, and the horizontal axis is $\theta (0,0)$, the alignment at $z = 0$. The dashed line in each violin plot represents the median, and the dotted lines represent 25 and 75 percentiles.
Though the median of $\Delta\theta$ at high-$\theta (0,0)$ seems to be slightly underside, a significant trend is not found because of the large dispersion. 

Although 1 Gyr is sufficiently smaller than the typical evolutionary timescale of voids as mentioned in \ref{ssec:mergertree}, the dispersion of $\Delta\theta$ is about 10~\%, which is not so small. If voids deform by tidal field, $\Delta \theta < 0$ is expected because tidal field and void should be aligned with time.
However, a considerable number of voids have a positive $\Delta\theta$. 
Moreover, this result hardly depends on $PR$. Even the voids {that} retain most of the particles do not show the correlation between tidal field and void orientation. This fact implies that not a few voids exist whose shape is strongly affected by various factors other than the tidal field. They are possibly the effect of finite resolution of simulation or other gravitational force components such as higher multipole components or angular components of gravitational force, for example. Keep it in mind, though, that this result does not mean that all the voids are independent of the tidal field, as there are plenty voids with $\Delta \theta < 0$.

\subsubsection{Time Evolution of Void Ellipticity}
\label{ssec:timeevolution_e}

\begin{figure}
  \centering
  \includegraphics[width=8cm]{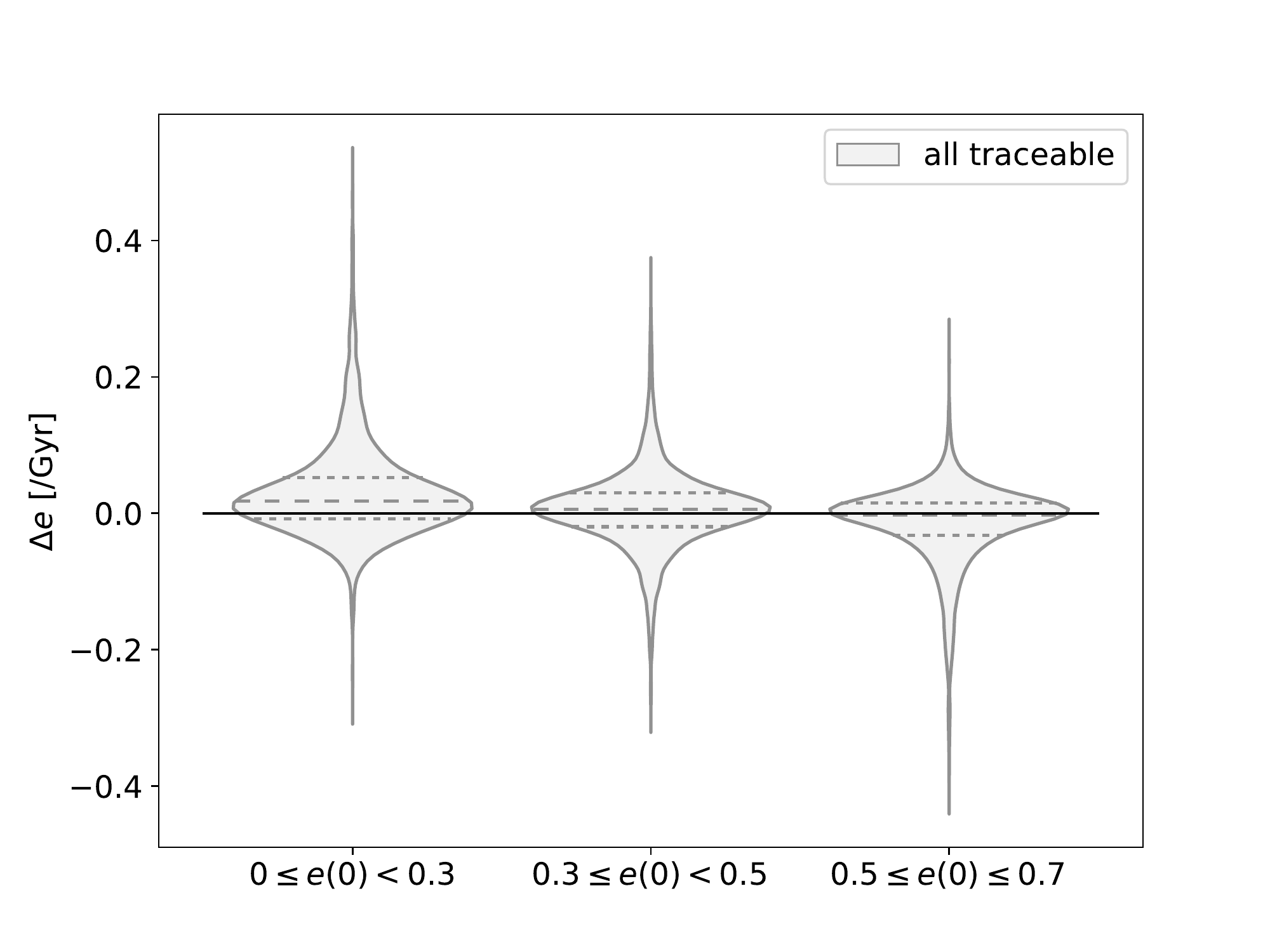}
  \caption{The distributions of ellipticity increase per Gyr at $z = 0$. The dashed line in each violin plot represents the median, and the dotted lines represent 25 and 75 percentiles.}
\label{fig: e_de_at}
\end{figure}

According to the results of {the} $N$-body simulation by \cite{Wojtak_2016}, the distorted void tends to become more spherical, and the spherical void tends to distort as it grows, although the shape of each void evolves variously. However, this tendency is not statistically recognized in our catalogue. Figure \ref{fig: e_de_at} shows the distribution of the increase of ellipticity in 1 Gyr at $z = 0$. The vertical axis of this figure is the increment per 1 Gyr of the ellipticity, and the horizontal axis is the ellipticity at $z = 0$. As in Figure \ref{fig: t_dt}, the dashed lines in each violin plot represent the median, and the dotted lines represent 25 and 75 percentiles. 
While the voids with small ellipticity seem to have a relatively large median of $\Delta e$, it is still not statistically significant because of the large dispersion as is the case for alignment.

One possible reason for the difference between \cite{Wojtak_2016} and our result may lie in the difference in the time scale. We focus on the short-term variation (differential quantity) at $z=0$ in 1 Gyr, while they focus on the variance from the early ($z=100$) void to the present ($z=0$) void. The larger the time interval, the larger $|\Delta e|$ is allowed and those with large $|\Delta e|$ can be strongly affected by the parameter space boundary; $\Delta e=1$ is only possible at $e=0$ and $\Delta e=-1$ is only possible at $e=1$. This boundary can cause a negative correlation between $e$ and $\Delta e$ even if $e$ is randomly changing.
Another possibility is the different definition of a void, but this would not be the main cause. They connect the neighbouring zones at $z=0$, which ensures the voids do not rapidly change their shape due to merger or segmentation. In contrast, we do not make zone connections stable as they do but can select voids that hardly experience merger and segregation by monitoring $PR$ parameter. 
Nevertheless, our result hardly depends on $ PR $ besides a slight change in the median of $\Delta e$. This fact implies that the difference in $e-\Delta e$ relation is hardly affected by whether we consider particle exchange or not.

\subsection{Correlation Between Tidal Field and Void Evolution}
\label{ssec:correlation}
If the tidal field is strong along the void major axis, the void will become more elongated. In this case, the{r}e should be a positive correlation between $T_\theta$ (the amplitude of the tidal field along void major axis) and $\Delta e$ (the ellipticity increment). To quantify the correlation between the $ T_\theta $ and $ \Delta e $, we calculate the correlation coefficient. The coefficient of $ T_\theta $ and $ \Delta e $ for the set of voids that satisfy condition $Z$ is given as
\begin{equation}
{\rm corrcoef}(T_\theta,\Delta e|Z)=\frac{{\rm Cov}(T_\theta, \Delta e|Z)}{\sqrt{{\rm Cov}(T_\theta, T_\theta|Z)\ {\rm Cov}(\Delta e, \Delta e|Z)}}, \label{eq_corrcoef}
\end{equation}
using components of covariance matrix:
\begin{align}
&{\rm Cov}(X,Y|Z)
=
\frac{1}{N_{Z}-1}
\sum_{i \in Z}^{N_{Z}}
\left(
X_i-\langle X \rangle_Z
\right)
\left(
Y_i-\langle Y \rangle_Z
\right),\\
&\langle X \rangle_Z
=
\frac{1}{N_{Z}}
\sum_{i \in Z}^{N_{Z}}
X_i
\end{align}
where the sum runs over voids {that} satisfy condition $Z$, and $N_{Z}$ is the number of such voids. $X, Y$ denote either $T_\theta$ or $\Delta e$.

\begin{figure}
  \centering
  \includegraphics[width=8cm]{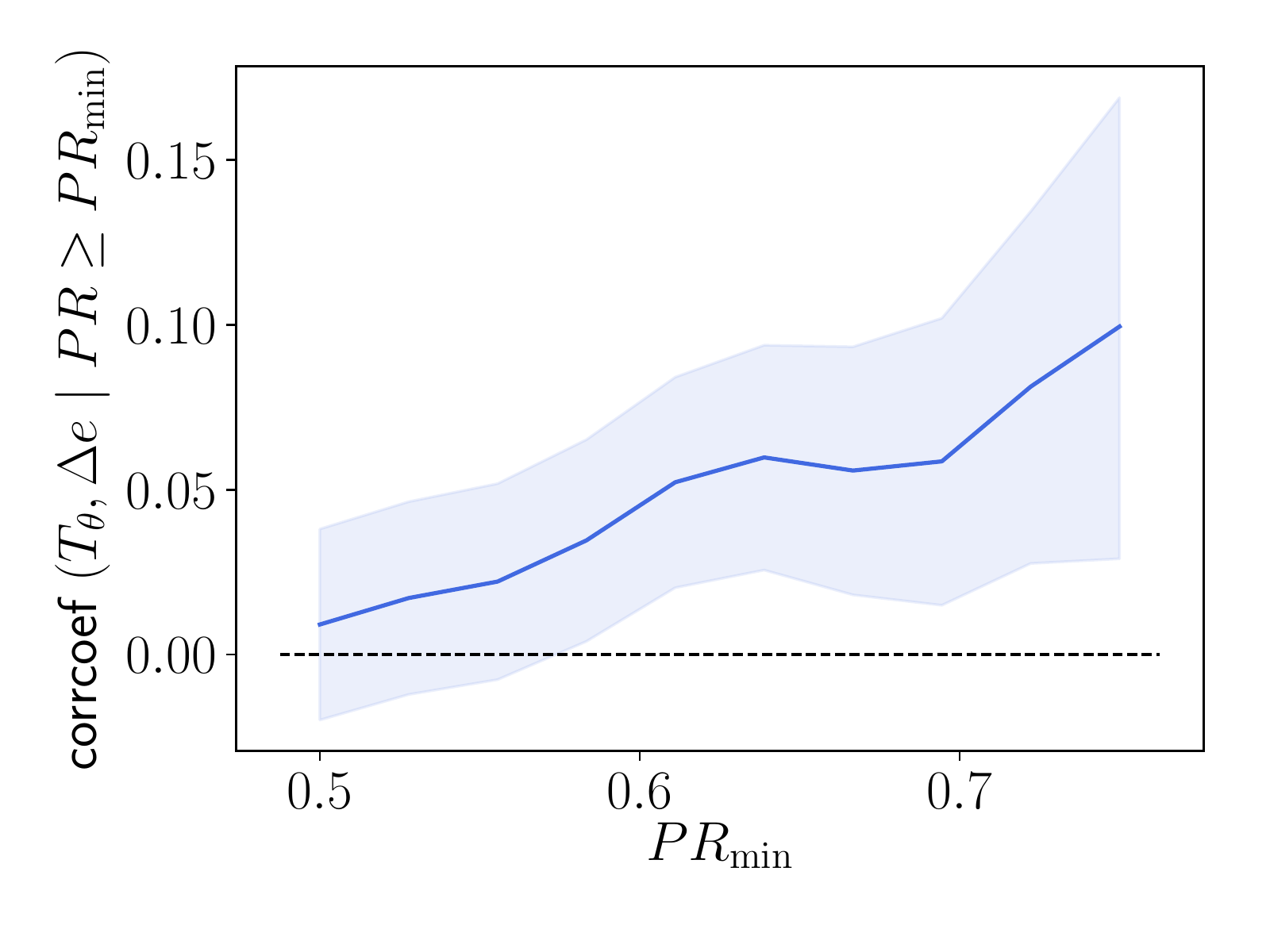}
  \caption{
  Correlation coefficients of $\Delta e$ and $T_\theta$ {that are} calculated by using the voids with particle retention rate $PR$ greater than or equal to a lower limit represented by $PR'$. The shaded region shows 95\% confidence interval.
  }
\label{fig: PR_Tde}
\end{figure}

The correlation coefficient between the tidal force $T_\theta$ and the ellipticity increase $\Delta e$ is shown in Figure \ref{fig: PR_Tde} as a function of {the} minimum value of particle retention parameter $PR$. The shaded region in the figure represents a 95~\% confidence interval calculated by equation \ref{eq_confcorr}. 
The voids whose major axes rotate over 45 degrees {are excluded here} because they are almost spherical and major and minor axes are easily interchanged in 1 Gyr. However, such voids comprise only about 5~\% of the total traceable voids, and the result hardly changes so much even if we include them.

\begin{figure}
  \centering
  \includegraphics[width=8cm]{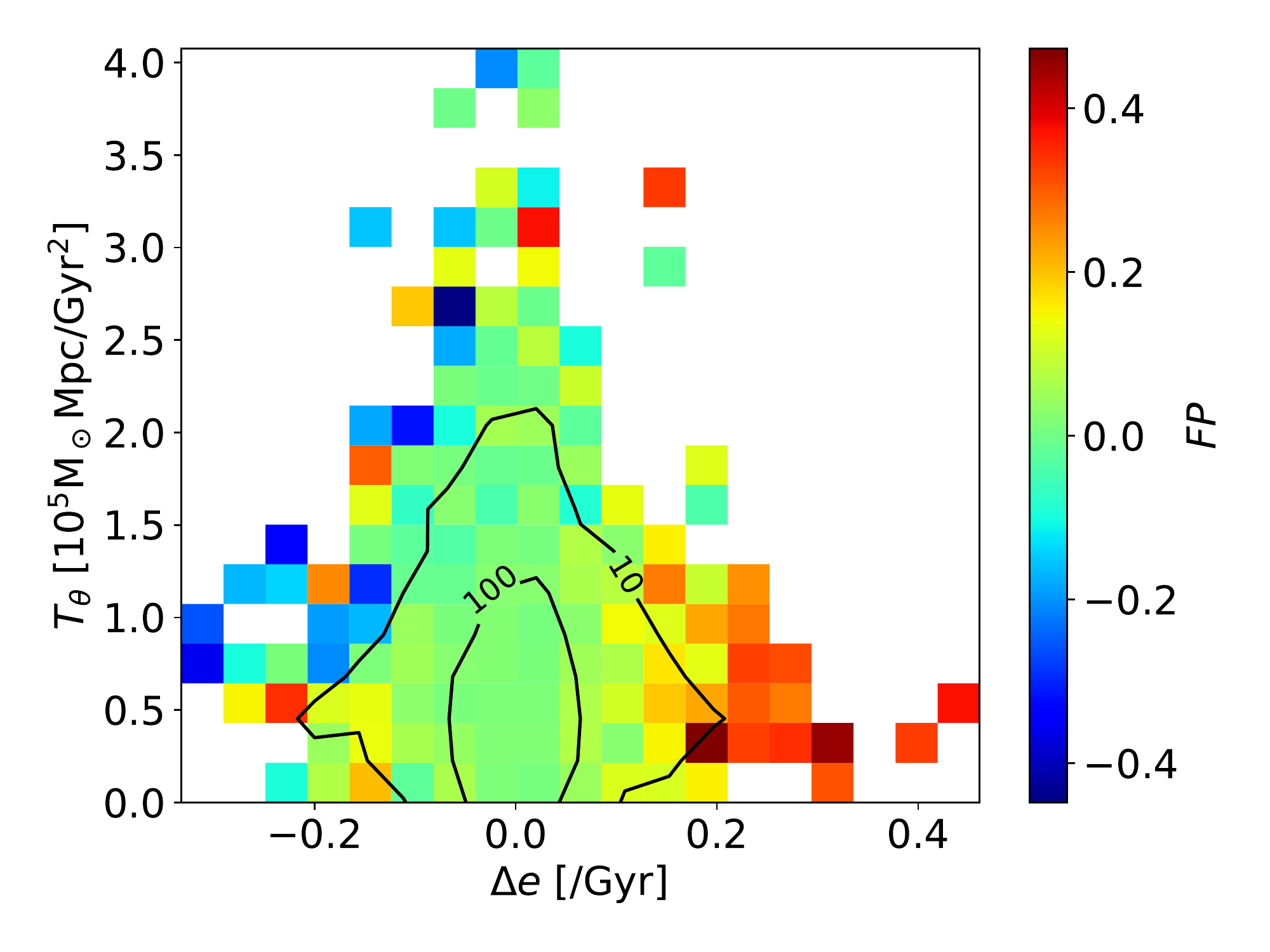}
  \caption{
  Tidal force $T_\theta$ against $\Delta e$. Black contours indicate void number count. The colour denotes $FP$, which means that if it is high, the void gains most of the particles owned by the void after evolution from outside of the void, and if it is low, it means that the void parts with most of its particles the void owned before evolution (see also section \ref{ssec:mergertree}).
  }
\label{fig: de_T}
\end{figure}

The correlation coefficient is zero-consistent within the 95~\% level when we include small $PR$ voids, $PR<0.5$, while it takes significant positive value if we limit the sample with $PR>0.6$.
This result indicates that the voids which retain particles before and after evolution are distorted by the tidal field. It is worth noting, however, that averaged overall $PR$, the correlation coefficient becomes consistent with zero with our definition of a void. 
We show the relation between the tidal force $T_\theta$ and the ellipticity increase $\Delta e$ in more detail in Figure \ref{fig: de_T}. Black contours in Figure \ref{fig: de_T} represent the number distribution of voids. The colours in Figure \ref{fig: de_T} represent $FP$, which reflects the amount of particle exchange. 
Quantity $FP$ indicates whether particles have entered or exited; if it is positive (negative), it means that the void has lost (gained) dark matter particles in 1 Gyr. Although $PR$ is also an index indicating whether or not particles are exchanged, we use $FP$ here to discriminate whether a void has gained or lost particles. 

The variance in $\Delta e$ is large where $T_\theta$ is small, {depicting that} the significant shape distortion occurs where the tidal field is relatively weak. It is expected that the external tidal field cannot be the main reason to distort the shape of the voids for voids with $T_\theta < 10^5 [M_\odot \mbox{Mpc}\,\mbox{Gyr}^{-2}]$. 
On the other hand, the tendency is especially prominent at the low-$T_\theta$ side in Figure \ref{fig: de_T} that the smaller the value on the horizontal axis, the smaller the value of $ FP $. Conversely, the larger the value on the horizontal axis, the larger the value of $ FP $. This fact means that a void tends to be distorted when it gets particles from its surroundings, and if it loses particles, it tends to become closer to a sphere. Deformation {owing} to this effect produces a large variance in the increase in void ellipticity $ \Delta e $, leading to a lower correlation between the shape evolution tendency of the void and the tidal field. 
As shown in Figure \ref{fig: PR}, voids with large $|FP|$ generally have small $ PR $. Therefore, if we select only high-$PR$ voids, most of the voids with large shape variance and small $T_\theta$ {are} removed. 
We repeat the analysis of Figure \ref{fig: PR_Tde} as the function of the minimum value of $|FP|$, instead of $PR$. 
However, the correlation is very weak compared with that shown in Figure \ref{fig: PR_Tde}. Therefore, we conclude that the $PR$ is more suitable for isolating the void population which is affected by the tidal force.

\subsection{Proxy of \texorpdfstring{$PR$}{PR}}
\label{ssec:proxy_of_PR}
\begin{figure}
  \centering
  \includegraphics[width=8cm]{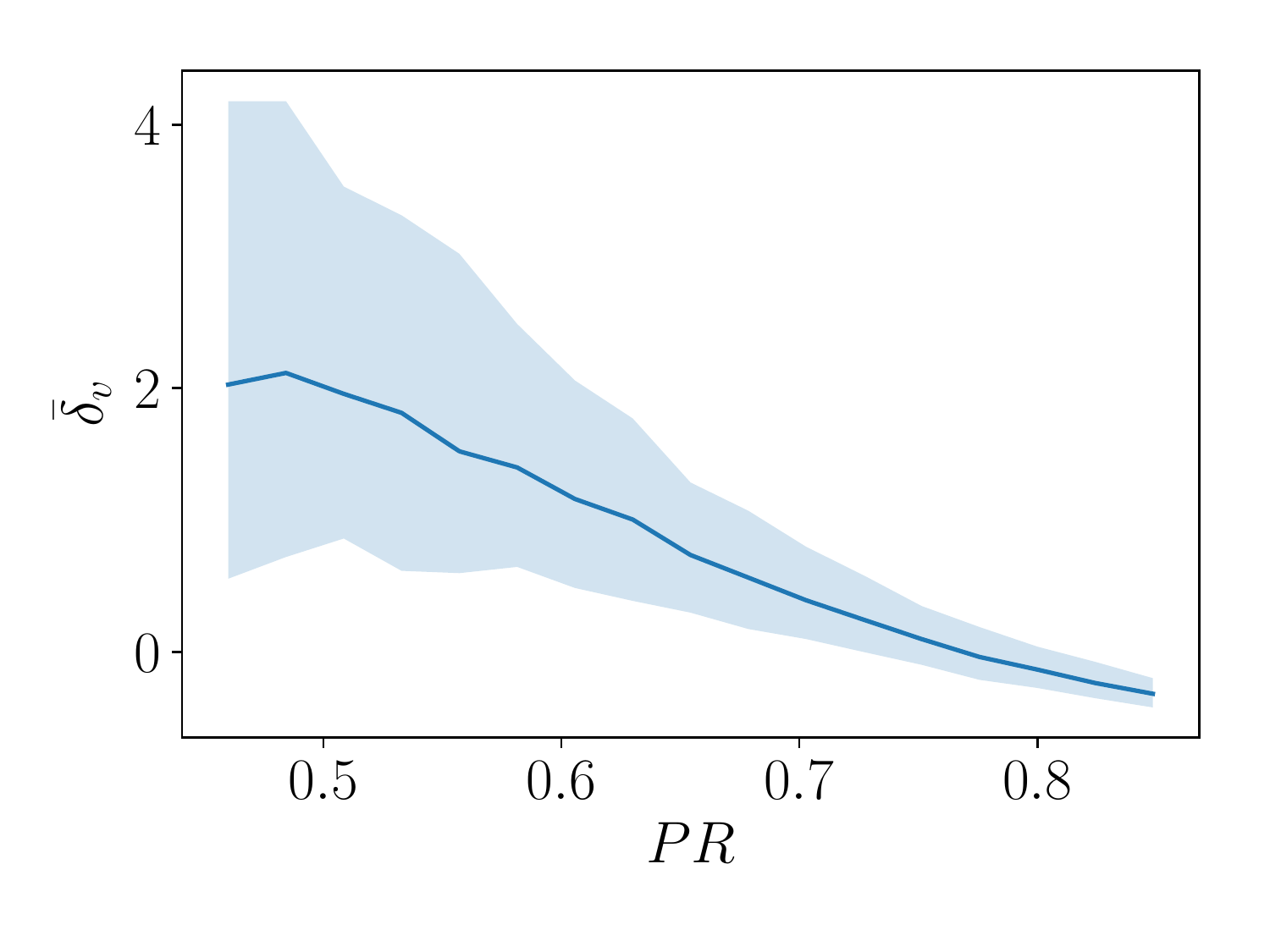}
  \caption{
  Mean overdensity in void against particle retention $ PR $ (see text for definition). The shaded region represents the standard deviation at each $ PR $ bin.
  }
\label{fig: PR_deltav}
\end{figure}

The particle retention parameter $PR$ cannot be determined from observation. However, we find that the average density in the void is strongly correlated with $ PR $. Figure \ref{fig: PR_deltav} shows the relation between the average density of a void $ \bar \delta_v $ and $ PR $. $ \bar \delta_v $ is defined as
\begin{equation}
    \bar\delta_v=\frac{\rho_v}{\bar\rho}-1,
\end{equation}
where $\rho_v$ is void mass density defined in \ref{ssec:voidfind} and $\bar\rho$ is the average mass density of the Universe. Hence, it can be effectively determined by the mass of the wall surrounding a void. In Figure~\ref{fig: PR_deltav}, the shaded region shows the standard deviation and the solid line indicates the average of $ \bar \delta_v $ in each $ PR $ bin. 
We find a clear anti-correlation between $\bar{\delta_v}$ and $PR$.

\begin{figure}
  \centering
  \includegraphics[width=8cm]{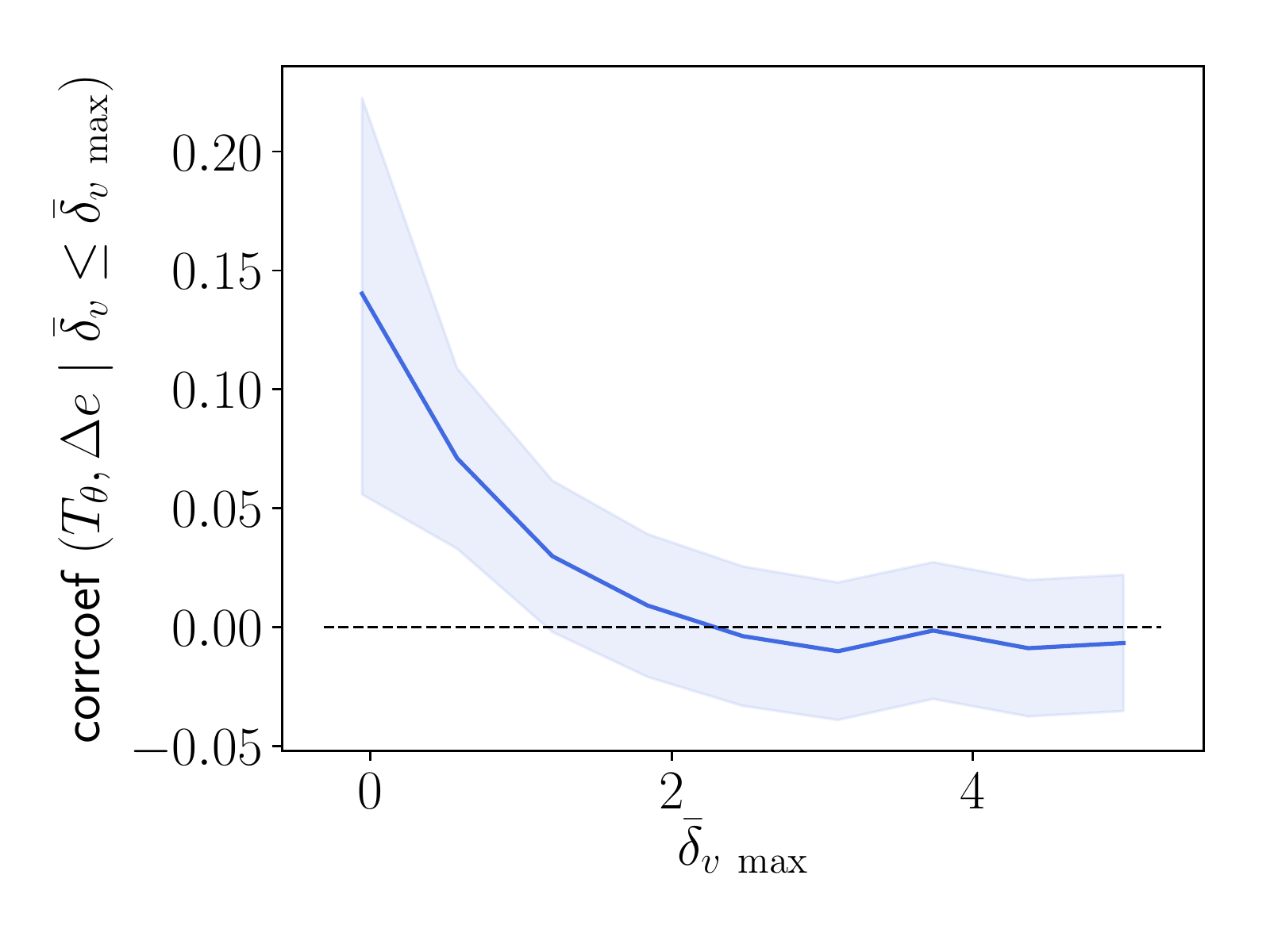}
  \caption{
  Correlation coefficient between $ T_\theta $ and $ \Delta e $ for the voids whose average over density $\bar\delta_v$ is less than or equal to arbitrary given $\bar\delta_v'$. The shaded region indicates the 95 \% confidence interval.
  }
\label{fig: deltav_corr}
\end{figure}

Using this relation, $ \bar \delta_v $ can be used as a proxy of $ PR $. We revisit the correlation coefficient analysis as a function of the maximum value of $\bar{\delta_v}$, instead of the minimum value of $PR$ in Figure \ref{fig: deltav_corr}. We find that the significant positive correlation between $T_\theta$ and $\Delta e$ appears with more than $95\%$ confidence when the upper limit of $\delta_v$ is less than around 1.

\begin{figure}
  \centering
  \includegraphics[width=8.5cm]{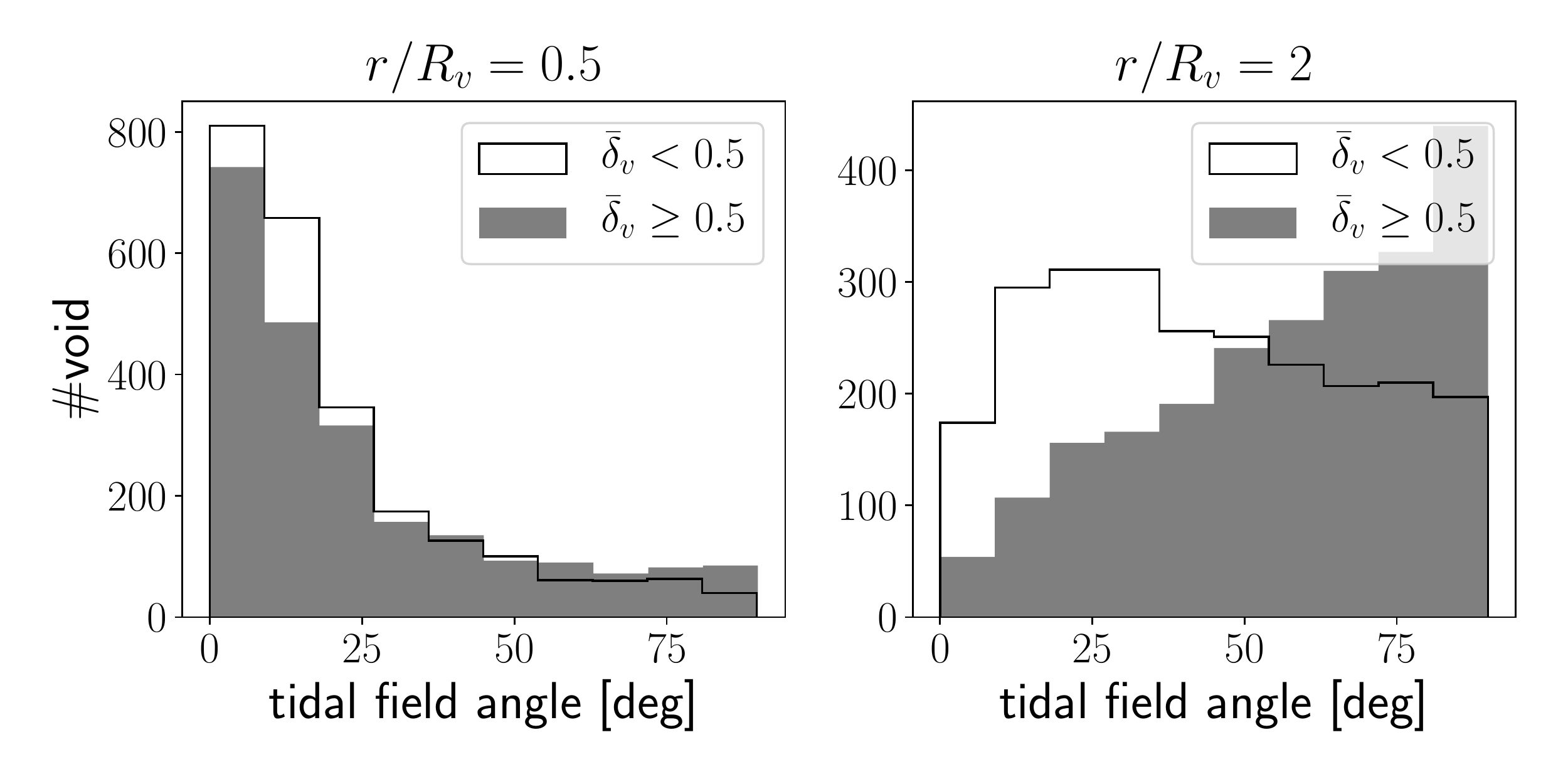}
  \caption{
  The distribution of the angle between $\hat{\boldsymbol n}_{\max}^{(2)}$ at $R_v$ and $r=0.5R_v$ (the left panel) or $r=2R_v$ (the right panel). The open histogram is for the voids with $\bar \delta_v<0.5$ and filled histogram is for the voids with $\bar \delta_v\ge 0.5$.
  }
\label{fig: r_theta}
\end{figure}

This relation between $ PR $ and $ \bar \delta_v $ implies that a void with higher mass density exchanges a larger proportion of particles, which can be explained {below}. For the voids with {a} larger $\bar{\delta_v}$, the density of the surrounding area is higher, and non-linearity becomes prominent.
Therefore, a tidal field around a void may mainly arise from the non-linearly evolving small-scale fluctuations, and the large-scale modes across the void hardly affect the tidal field around the void. If the small scale structures dominate, the tidal field in the radial direction can fluctuate on smaller scales than the size of the void.

To confirm this, the {inner (outer)} tidal fields measured on a sphere with a radius of half (twice) of $R_v$ {are compared} with the one measured on $R_v$. Figure \ref{fig: r_theta} shows the distributions of the angle between the direction of the tidal field at $R_v$ and 0.5 $R_v$ (left) and 2$R_v$ (right). Voids are {classified} into two groups; the voids with $\bar \delta_v<0.5$ (open-histogram) and $\bar \delta_v\ge 0.5$ (filled-histogram). {It is noted} that the distributions in this figure are not corrected for the volume effect (correspond to $P(\theta)$ in equation \eqref{P_theta}) and thus they follow sine function iftwo random directions {make the angle}. 
In the left panel, the tidal field inside the void is aligned to the tidal field at $r=R_v$ in most cases. On the other hand, in the right panel, the voids with low $\bar \delta_v$ tend to have a smaller variation of tidal field direction than the voids with high $\bar \delta_v$. The high-density voids seem to have the tidal field outside the void which faces almost independent direction to {that of} $r=R_v$. This fact means that the quadrupole component of the gravitational field fluctuates on a short scale near a void especially in the case that the void has {a} high $ \bar \delta_v $. 
Such a small-scale fluctuation in the gravitational field can cause particle exchanges, and the tidal field surrounding the void is no more enough to explain the shape evolution of the void. However, for low $ \bar \delta_v $ voids, the tidal field is relatively coherent up to the outside of the void, and the description that a void evolves by background gravitational field seems to be well supported.

\section{Summary}
\label{sec:summary}

We have investigated the correlation between the shape evolution of cosmic voids and the tidal field around them at around $z=0$ by $N$-body simulations. 
As is well known, the shape of a void in the cosmic web is distorted (and become{s} more and more distorted) on average, which is confirmed in section \ref{ssec:voidcatalogue}. However, it is not {evident} since \cite{Icke_1984} has found that the low-density region itself approaches a sphere. {Although} it is expected that the tidal field distorts the shape of a void, it {is} still unclear whether all the voids are affected by the tidal field in the same manner. 

Tracing individual voids, we have found out {that} the voids are full of individuality and change its shape by the amplitude of the surrounding tidal field {and} particle exchange. 

{The} results on the evolution of individual voids in 1 Gyr and the tidal field on the void scale {are summarized below}:
\begin{enumerate}
    \item We do not find a significant tendency in the evolution trend of ellipticities and alignments, {owing} to the very large intrinsic scatter.
    \item A positive correlation between $T_\theta$ (vector component of tidal force in the direction of the void principal axis) and an increase in ellipticity is found only for voids with little particle exchange.
    \item A negative correlation between particle retention and average void density exists. {A} positive correlation {appears} again as with voids with high particle retention{, on} examining the correlation between $T_\theta$ and ellipticity increment for low-density voids.
    \item High-density structures around a void shorten the coherent scale of the surrounding tidal field, which can be a cause of particle exchanges.
\end{enumerate}
{The second point} (ii) suggests that if $PR$ is high, the shape of a void evolves with reflecting the tidal field. {An investigation of} the voids with low particle retention shows that voids tend to be distorted when the particles are obtained and rounded when the particles are lost. This causes a large variance in the time evolution of the ellipticity and hides sign of the response to the quadrupole component of the gravitational field at void scale, as discussed in section \ref{ssec:timeevolution_e}. Such voids tend to have a higher average density. When the average void density ($ \approx $ density of wall around the void) is large, the quadrupole component of radial gravity turns significantly around the void, and it can be {one of the reasons} why particle exchange often occurs in a high-density void. Conversely, voids {with} a very low average density, have a positive correlation between the effective tidal field and the increase in ellipticity, which is a sign of pure dynamical evolution by the tidal force.

\section*{Acknowledgements}

This work is partially supported by MEXT KAKENHI Grant-in-Aid for Scientific Research on Innovative Areas
'Cosmic Acceleration' (15H05890 and 16H01096 (AN)) and by JSPS Grants-in-Aid for Scientific Research (17H01110 and 19H05076).
This work has also been supported in part by the Sumitomo Foundation Fiscal 2018 Grant for Basic Science Research Projects (180923), and the research collaboration funding of the Institute of Statistical Mathematics 'New Development of the Studies on Galaxy Evolution with a Method of Data Science'.
{The authors would like to thank Enago (www.enago.jp) for the English language review.}

\section*{Data Availability}

The data that support the findings of this study will be shared on reasonable request to the corresponding authors.

\bibliographystyle{mnras}
\bibliography{ms}

\appendix
\section{Confidence interval for correlation coefficients}
We derive confidence interval of the correlation coefficient given by equation \eqref{eq_corrcoef} referencing \cite{Anderson_1958}. Here, the sample correlation coefficient is written as $ c $ and the population correlation coefficient is written as $ c_{g} $. Using Fisher's z-transformation (inverse hyperbolic tangent function)
\begin{equation}
z(x)=\frac{1}{2}\log_{e}\frac{1+x}{1-x} \label{eq_z},
\end{equation}
it is known that $ z(c) $ is normally distributed around $ z (c_{g}) $ with variance $ 1 / (n-3) $ when the number of samples $ n $ is large enough \citep{Fisher_1915}.
That is, $ z_{\rm norm} = (z(c) - z(c_{g})) / (1 / \sqrt{n-3}) $ has a normal distribution with mean 0 and variance 1.

With this fact, the $ p $ \% confidence interval of a given correlation coefficient can be calculated as below. First, the top ((100 - p) / 2) \% percentile of the standard normal distribution is given by
\begin{equation}
P(p)=\sqrt{2}\ {\rm erfc}^{-1}(1-p/100),
\end{equation}
where
\begin{equation}
{\rm erfc}(x)=\frac{2}{\sqrt{\pi}}\int^{\infty}_{x}e^{-t^2}dt
\end{equation}
is the complementary error function, which satisfies the relation $ {\rm erfc} (x) = 1-{\rm erf} (x) $ with the error function 
\begin{equation}
{\rm erf}(x)=\frac{2}{\sqrt{\pi}}\int^{x}_{0}e^{-t^2}dt.
\end{equation}
Therefore, assuming that $ z_{\rm norm} $ is between $ -P(p) $ and $ P(p) $, the section where $ z(c_g) $ exists with the probability of p\% can be obtained; postulating
\begin{equation}
-P(p)\le\frac{z(c)-z(c_g)}{1/\sqrt{n-3}}\le P(p),
\end{equation}
we obtain the range of $z(c_g)$ as below:
\begin{equation}
z(c)-\sqrt{n-3} P(p)\le z(c_g)\le z(c)+\sqrt{n-3} P(p).
\end{equation}
Performing the inverse transformation of equation \ref{eq_z}, this inequality is transformed as
\begin{equation}
z^{-1}\left(z(c)-\sqrt{n-3} P(p)\right)\le c_g\le z^{-1}\left((z(c)+\sqrt{n-3} P(p)\right).
\end{equation}
Since the transformation z is arctanh, we finally obtain
\begin{equation}
\begin{split}
&\frac{f_{-}-1}{f_{-}+1} \le c_g \le \frac{f_{+}-1}{f_{+}+1},\\
&f_{\pm}=\exp\left[2\left(\frac{1}{2}\log_{e}\left(\frac{1+c}{1-c}\right)-\frac{P(p)}{\sqrt{n-3}}\right)\right]  \label{eq_confcorr}
\end{split}
\end{equation}
as the p\% confidence interval of the correlation coefficient.


\bsp    
\label{lastpage}
\end{document}